\documentclass[twocolumn,amsmath,amssymb,prb,longbibliography]{revtex4-2}
\usepackage{multirow}
\usepackage{array}
\usepackage{amsmath}
\usepackage{graphicx}% Include figure files
\usepackage{dcolumn}% Align table columns on decimal point
\usepackage{bm}% bold math
\usepackage{verbatim}
\DeclareMathAlphabet \mathbfcal{OMS}{cmsy}{b}{n}

\begin{document}

\title{Ultrafast Field-driven Valley Polarization of Transition Metal Dichalcogenide Quantum Dots}
\author{Aranyo Mitra}
\email{amitra1@gsu.edu}
\author{Ahmal Jawad Zafar}
\author{Vadym Apalkov}
\affiliation{Department of Physics and Astronomy, Georgia State University, Atlanta, Georgia 30303, USA}
\begin{abstract}
We study theoretically the electron dynamics of transition metal dichalcogenide quantum dots in the field of an ultrashort and ultrafast circularly polarized optical pulse. The quantum dots have the shape of a disk and their electron systems are described within an effective model with infinite mass boundary conditions. Similar to transition metal dichalcogenide monolayers, a circularly polarized pulse generates ultrafast valley polarization of such quantum dots. The dependence of the valley polarization on the size of the dot is sensitive to the dot material and, for different materials, show both monotonic increase with the dot radius and nonmonotonic behavior with a local maximum at a finite dot radius.

%The pulse applied has a duration of a few femtoseconds, and electron dynamics is found to be coherent in the presence of this field. We apply here a circularly polarized pulse and study the valley polarization generated after it. Valley polarization is defined as the normalized difference in the populations of the conduction band in two valleys – $K$ and $K’$. The valley polarization is found to be size-dependent and increases as the radius is increased. Comparison between different transition metal dichalcogenide materials shows enhanced degrees of generated valley polarization as a function of size and field amplitude $F_0$ of the pulse, and dependent on the type of material. A significantly larger valley polarization is generated, compared to previously reported values for monolayers, for relatively small values of the Field amplitude $F_o$ ($\leq$  0.15 V/nm).
\end{abstract}
\maketitle
\section{INTRODUCTION}
In recent years, atomically thin transition metal dichalcogenide (TMDC) systems have garnered attention and been the subject of extensive studies \cite{RN235,RN217,RN236}. The unique optical, topological and transport properties,\cite{RN259,RN260} as well as a direct band gap in the visible frequency range \cite{RN217,RN218,RN228} and a substantial carrier mobility \cite{RN221} of this class of materials make them suitable candidates for application to optoelectronic, spintronic and valleytronic applications.\cite{RN234,RN258}

TMDC monolayers are made of a honeycomb-lattice structure consisting of a group-VIB transition metal atom, coordinated by six chalcogen atoms in a trigonal prismatic configuration \cite{RN231,RN218}. The presence of different atoms in the structure leads to an inversion ($\mathcal{P}$) asymmetry of the TMDC monolayers and opening of band gaps at the 
$K$ and $K'$ valleys - two degenerate, inequivalent valleys in the reciprocal space, with the degeneracy being protected by the time-reversal ($\mathcal{T}$) symmetry \cite{RN232}. At these valleys, electrons posses nontrivial topological properties. The corresponding Chern numbers, i.e., topological charges, at the $K$ and $K'$ valleys are equal in magnitude and have opposite sign, giving rise to a net zero topological charge for the total Brillouin zone. \cite{RN271}
In TMDCs, the inversion asymmetry and the topological effects at individual valleys, along with a strong spin-orbit coupling leads to circular dichroism manifested in the form of optical selection rules for interband transitions, at high symmetry points in the reciprocal space, notably at the $K$ and $K’$ valleys \cite{RN203,RN230,RN224}. These optical selection rules  also depend on the spin degree of freedom and enable optical excitations of different helicities and frequencies to excite carriers with different spin and valley index combinations\cite{PhysRevLett.108.196802}, which has potential applications in integrated valleytronics and spintronics.
As a result, a circularly polarized light is mainly coupled to one of the valleys only, i.e., to $K$ or $K'$ valley, depending on the helicity of the pulse. This preferential coupling leads to an asymmetry in the generated carrier populations of the $K$ and $K’$ valley after interaction with a circularly polarized optical pulse resulting in a finite valley polarization of the system.  Thus, the valley degree of freedom can be controlled on a femtosecond timescale if the incident optical pulse is ultrashort with a few oscillations of the field. The interaction of a TMDC monolayer with a circularly polarized ultrashort and ultrastrong optical pulse was studied theoretically in Ref.\cite{RN230}, where a large residual valley polarization induced by an ultrashort pulse was reported. The generation of such a large valley polarization in 2D TMDC monolayers is due to the effect of topological resonance, when the dynamical phase and the topological phase, which is gradually accumulated during the pulse, compete with each other.\cite{RN227}

The band-edge dispersion in TMDC monolayers is well described by a massive Dirac model with strong coupling between the spin and valley degrees of freedom. As a consequence of such a strong valley-spin coupling, a pronounced spin-splitting of the valence band was observed \cite{RN269}, and together with a large separation of the two valleys in the reciprocal space, it protects against intervalley scattering, as discussed in Refs.\cite{RN109,RN222,RN229,RN223}.

The ultrafast valley polarization of TMDC monolayers can be controlled by the amplitude of a pulse and by the type of TMDC material. 
Here we propose another  method of control of the generated ultrafast valley polarization in TMDC systems. Namely, we consider nanoflakes or quantum dots (QDs) of TMDC monolayer. 
In QDs, the translational symmetry is broken and the carriers are confined in all spatial directions and occupy spectrally sharp energy levels akin to atomic energy levels \cite{RN240}. Quantum dots, sometimes referred to as artificial atoms, have specific properties, which are directly related to their zero-dimensional nature: the Hund's rule \cite{RN261}, which determines the population of QD energy levels; the Kondo effect \cite{RN263}; and the Coulomb blockade \cite{RN262}, which is due to electron-electron interactions, and so  on. 
 
The discrete nature of the energy spectra of QDs depends on the dots' size and their geometry. Thus, the interaction of such QDs with an ultrashort optical pulse should be sensitive to the QDs' geometric parameters. In this case, the valley polarization, induced by an ultrashort circularly polarized pulse, should depend on the size of a nanoflake. Below we study such dependence for four different TMDC materials: MoS\textsubscript{2}, MoSe\textsubscript{2}, WS\textsubscript{2} and WSe\textsubscript{2} \cite{RN239}. 
The QDs considered below are of the size of a few nanometers, which facilitates the control and optimization of QD-based nanoscale devices, since the reduced size enables the QDs to be populated by a few electrons. These nanoscale systems also possess useful optical and transport properties, much like their monolayer counterparts, with applications across different scientific areas, particularly in quantum computing \cite{RN266}, semiconducting laser systems \cite{RN265} and biomedical applications \cite{RN264}.\\
%table for parameters
\begin{table}[ht]
\centering % used for centering table
\begin{tabular}{c c c c c} % centered columns (5 columns)
\hline\hline %inserts double horizontal lines
Material & $a$ & $t$ & $\Delta$ & $\lambda_{so}$  \\ % inserts table
%heading
\hline % inserts single horizontal line
MoS\textsubscript{2} & 3.193 & 1.10 & 1.66 & 0.075 \\
        MoSe\textsubscript{2} & 3.313 & 0.94 & 1.47 & 0.09\\
        WS\textsubscript{2} & 3.197 & 1.37 & 1.79 & 0.215\\
        WSe\textsubscript{2} & 3.310 & 1.19 & 1.60 &0.23\\ % [1ex] adds vertical space
\hline %inserts single line
\end{tabular}
\label{table:nonlin} % is used to refer this table in the text
    \caption{Parameters for different TMDC materials, as reported in Ref. \cite{PhysRevLett.108.196802} using first-principles band structure calculations. The unit for $a$ is \AA, and for $t, \Delta$ and $\lambda_{so}$ is $eV$}
    \label{tab1}
\end{table}

In the present paper, we consider TMDC QDs in the shape of a disk of radius $R$, which determine the QD size and is considered as a parameter that can be used to tune the valley polarization of the system. 
The electron dynamics of such QDs is described within the two-band effective \textbf{\textit{k.p}} model \cite{PhysRevLett.108.196802,RN237,RN238}, which is applied separately to the $K$ and $K'$ valleys of TMDC QD system. 
The model is constructed using symmetry considerations, based on the band structure of the TMDC system. The conduction and valence band wave functions within this model are related to the $d$ bands of the transition metal\cite{PhysRevLett.108.196802}. The interaction of an ultrashort circularly polarized optical pulse with TMDC QDs is expected to generate the residual, i.e., after the pulse, conduction band population, which is different for two TMDC valleys, thus generating a residual valley polarization of the system.

This paper is organized as follows. In Sec. \ref{sec2}, we introduce the theoretical model and the main equations. We also define TMDC QDs the shape of the optical pulse considered in our calculations. In Sec. \ref{sec3}, we present the numerical results obtained for different sizes of TMDC QDs and for different TMDC materials. We summarize our concluding remarks in Sec. \ref{sec4}.

\section{MODEL AND MAIN EQUATIONS}\label{sec2}

We consider a TMDC quantum dot in the shape of a disk with radius $R$. The TMDC system has two valleys, $K$ and $K'$, which we label below with index $\zeta $, where $\zeta = 1$ for valley $K$ and $\zeta = -1$ for valley $K'$. For each valley, the QD system is described within a low-energy effective model Hamiltonian, \cite{PhysRevLett.108.196802} which has the following form:
\begin{equation}\label{eqn1}
    \mathcal{\widehat{H}}_{\zeta }= \mathcal{\hat{H}}_o +\frac{\Delta}{2}\hat{\sigma}_z  - \lambda_{so}\zeta \frac{\hat{\sigma}_z - 1}{2}\hat{s_z} + V(\mathbf{r})\hat{\sigma_z},
\end{equation}
where $\mathcal{\widehat{H}}_o=at(\zeta k_x\hat{\sigma_x}+k_y\hat{\sigma_y} )$ is the massive Dirac-type two band \textit{\textbf{k.p}} Hamiltonian, $\textbf{k} = - \iota\nabla$, $a$ is the lattice constant, $t$ is the effective hopping integral, $\Delta$ determines the band gap of the material, $\lambda_{so}$ is the spin-orbit coupling constant that quantifies the valence band spin-splitting, and $s_z = \pm 1$ corresponds to spin-up ($+1$) and spin-down ($-1$) states.
Here, $\hat{s_z}$ and $\hat{\sigma}_{x,y,z}$ are the Pauli matrices acting on the spin and orbital degrees of freedom, respectively. The confinement potential, $V(\textbf{r})$, is defined in the mass form as $V(\textbf{r})= \Delta(\textbf{r}) \sigma_{z}$, where $\Delta(\textbf{r})$ is zero inside the quantum dot, i.e., $r<R$, and $\Delta(\textbf{r}) \to \infty$ outside. Such a profile of the confinement potential corresponds to infinite mass boundary conditions at the quantum dot boundary \cite{RN213,RN268,RN270}, i.e., at $r=R$. 
Although the properties of TMDC quantum dots dependent on the type of the edges, e.g., armchair or zigzag edges, the infinite mass boundary condition is a good approximation for TMDC nanostructures synthesized through lithographic techniques \cite{RN267}.
The parameters of the model, $a$, $t$, $\Delta$, and $\lambda_{so}$, are shown in Table \ref{tab1} for different TMDC materials, which are used below in our calculations.

%Add a sentence about the edges here
Withing the QD region, $r<R$, the effective Hamiltonian (\ref{eqn1}) takes the following form
\begin{equation}\label{eqn2}
\mathcal{\widehat{H}}_{\zeta } =
\begin{pmatrix}
\Delta/2 & \hbar v_f (\zeta k_x - \iota k_y)\\
\hbar v_f (\zeta k_x + \iota k_y)& - \Delta/2+ \lambda_{so} \zeta s_z
\end{pmatrix}.
\end{equation}
Since the QDs considered below have cylindrical symmetry, the 
electron states can be characterized by a magnetic quantum number, $m$, which takes integer values, $m= 0,\pm 1, \pm 2, ...$. Then, in the polar coordinates, $(r,\theta)$, the valley-dependent eigenfunctions of Hamiltonian (\ref{eqn2}) take the following form
\begin{equation}\label{eqn3}
    \Psi^{(\zeta)}(r,\theta) =
    \begin{pmatrix}
\psi_1^{(\zeta)} (r, \theta) \\
\psi_2^{(\zeta) (r, \theta) }
\end{pmatrix} =
    e^{im\theta}
\begin{pmatrix}
\chi_1^{(\zeta)}(r)\\
\chi_2^{(\zeta)}(r) e^{i \zeta \theta}
\end{pmatrix} 
\end{equation}
The functions $\chi_1^{(\zeta)}(r)$ and $\chi_2^{(\zeta)}(r)$ satisfy the following system of differential equations
\begin{equation}\label{eqn4}
\frac{\delta}{2} \chi_1^{(\zeta)} - i \zeta 
\left( \mathbf{\nabla}_r + \frac{\zeta (m+\zeta)}{r} \right) \chi_2^{(\zeta)}  =  \varepsilon^{(\zeta)} \chi_1^{(\zeta)}
\end{equation}
\begin{equation}\label{eqn5}
     -i \zeta \left( \mathbf{\nabla}_r - \frac{ \zeta m}{r} \right) \chi_1^{(\zeta)} + \left( \Lambda_{so} \zeta  s_z - \frac{\delta}{2}\right) \chi_2^{(\zeta)}  =  \varepsilon^{(\zeta)} \chi_2^{(\zeta)}
\end{equation}
where $\varepsilon^{(\zeta)} \equiv \frac{E}{at}$ , $\delta \equiv \frac{\Delta}{at}$, 
and $\Lambda \equiv \frac{\lambda_{so}}{at}$.
From Eq.~(\ref{eqn4}), we express $\chi_1^{(\zeta)}(r)$ in terms  of $\chi_2^{(\zeta)}(r)$ as
\begin{equation}\label{eqn6}
\chi_1^{(\zeta)}(r) =\displaystyle \frac{-i \bigl(\zeta \mathbf{\nabla}_r - \displaystyle\frac{(m+\zeta)}{r}\bigr)}{\bigl(\varepsilon^{(\zeta)} - \displaystyle \frac{\delta}{2}\bigr)}\chi_2^{(\zeta)}(r)
\end{equation}

%%%%%%%%%%%%%%%%%%%%%%%%%%%%%%%%%%%%%%%%%%%%%%%%%%
\begin{figure}[t]
    \centering
    \includegraphics[width=8.6 cm]{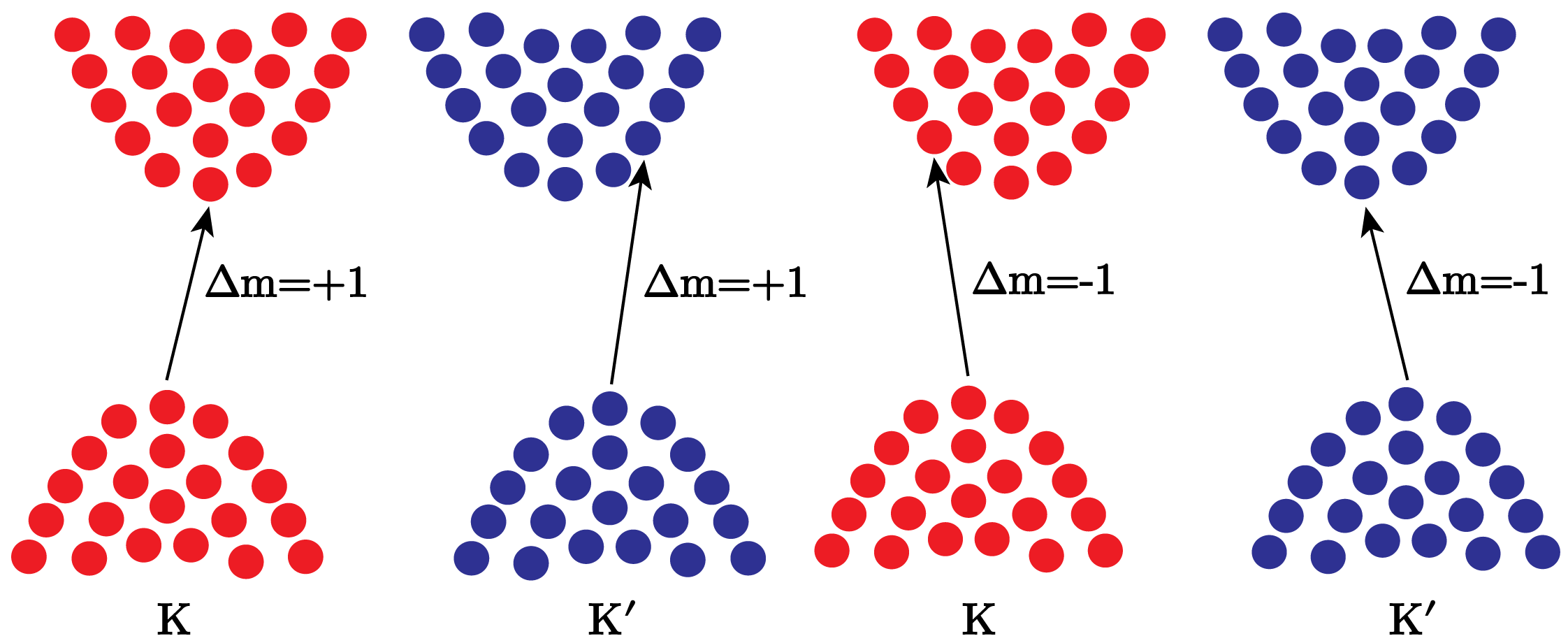}
    \caption{Illustration of the valley dependent optical selection rules and the corresponding transitions in each valley. For both conditions, $\Delta m=+1$ and $\Delta m=-1$, the asymmetry in populations is determined by the helicity of a pulse.  }
    \label{fig1}
\end{figure}

Then, substituting $\chi_1^{(\zeta)}(r)$ from Eq. (\ref{eqn6}) into Eq. (\ref{eqn5}), we obtain the following differential equation for $\chi_2^{\zeta}(r)$
\begin{equation}\label{eqn7}
    \nabla_r^{2}\chi_2^{\zeta}(r)+\frac{\mathbf{\nabla}_r}{r}\chi_2^{\zeta}(r)-\biggl( \frac{(m+\zeta)^{2}}{r^2}-\xi_{\zeta}^{2}\biggr) \chi_2^{\zeta}(r)=0
\end{equation}
Equation (\ref{eqn7}) is the Bessel equation, solution of which that is finite at the origin, $r=0$, has the following form 
\begin{equation}\label{eqn8}
    \chi_2^{(\zeta)}(r)= A^{(\zeta)}J_{|m+\zeta|}(\xi_{\zeta}r)
\end{equation}
where $J_{\nu}(x)$ is the Bessel function of the first kind of order $\nu$, $A^{(\zeta)}$ is the valley-dependent normalization constant, and $\xi_{\zeta} \equiv \sqrt{(\varepsilon^{(\zeta)} - \frac{\delta}{2})(\varepsilon^{(\zeta)} + \frac{\delta}{2}-\Lambda_{so} s_z)} $.  
Then, the expression for $\chi_1^{(\zeta)}(r)$ can be obtained from Eq. (\ref{eqn6}), see Appendix (\ref{app1pt1}), 
\begin{equation}\label{eqn9}
\chi_1^{(\zeta)}(r)= i \,A^{(\zeta)}\, \eta_{m}^{\zeta} \, \xi_{\zeta}\, \frac{J_{|m|}(\xi_{\zeta}r)}{\bigl(\varepsilon^{(\zeta)}-\frac{\delta}{2}\bigr)} ,
\end{equation}
where,
\begin{equation}\label{eqn10}
\eta_{m}^{\zeta} =
\begin{cases}
    -1 & (m+\zeta)>0\\
    +1 & (m+\zeta)<0\\
    \zeta & (m+\zeta=0)
\end{cases} .   
\end{equation}
%For the special case with values of $m=-\zeta$, $\chi_1^{(\zeta)}(r)=$.
%The general form of the eigenfunction $\Psi$ then becomes,
%\begin{equation}
%\end{equation}
\begin{figure}[t]
    \centering
    \includegraphics[width=8.6 cm]{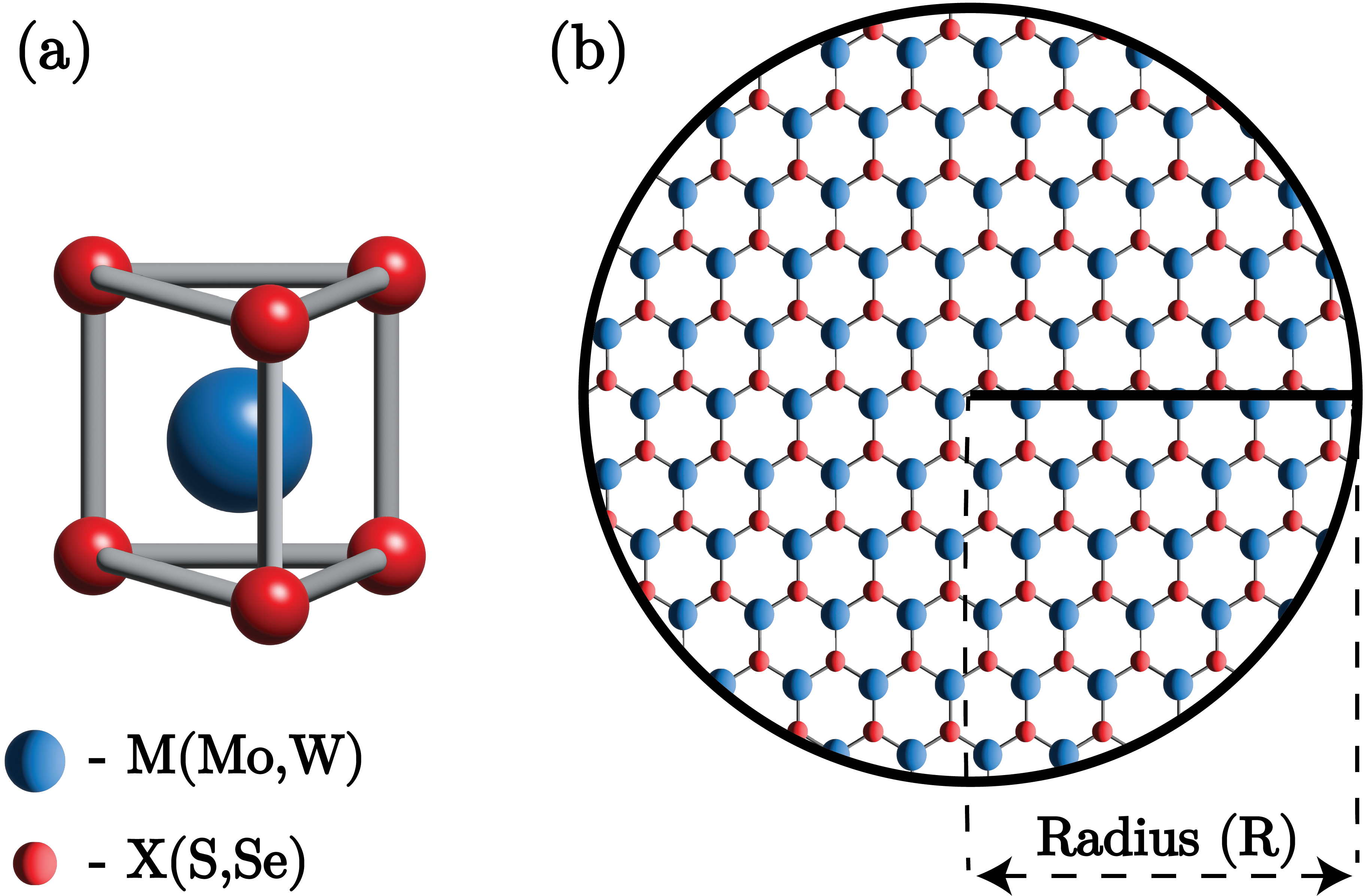}
    \caption{(a) Unit cell of $MoS_{2}$ monolayer. (b) Transition metal dichalcogenide quantum dot of radius $R$. }
    \label{fig2}
\end{figure}

The energy spectrum of a QD is determined from the infinite mass boundary condition\cite{RN213},
\begin{equation}\label{eqn11}
    \lim_{r\to \textbf{R}} \chi_2^{(\zeta)}(r) = i \zeta \lim_{r\to R} \chi_1^{(\zeta)}(r),
\end{equation}
where $R$ is the radius of the QD. Substituting Eqs. (\ref{eqn8}) and (\ref{eqn9}) into Eq. (\ref{eqn11}), we obtain the following equation for the eigen-energies of the QD system
\begin{equation}\label{eqn12}
        J_{|m+\zeta|}(\xi_{\zeta}r) = -\frac{\zeta \, \eta_{m}^{\zeta} \,\xi_{\zeta} \,J_{|m|}(\xi_{\zeta}r)}{(\varepsilon^{(\zeta)} - \frac{\delta}{2})}
\end{equation}
The dipole matrix elements between states $i$ and $j$ can then be calculated using the known eignefunctions $\Psi^{\zeta}(r,\theta)$ as:
\begin{equation}\label{eqn13}
    \textbf{D}_{ij}^{\zeta}=< \Psi^{\zeta}_{i} | e\textbf{r} |\Psi^{\zeta}_{j} >
\end{equation}
Substituting the eigenfunctions from Eq.(\ref{eqn3}) into Eq. (\ref{eqn10}), we obtain the $x$ and $y$ components of the dipole matrix elements in the following form
\begin{equation}\label{eqn14}
    D_{x,ij}^{(\zeta)}= e\Omega_x \int_{0}^{R} [\chi_{1,i}^{(\zeta) *}(r)\chi_{1,j}^{(\zeta)}(r)+ \chi_{2,i}^{(\zeta) *}(r)\chi_{2,j}^{(\zeta)}(r)] r^2 \,dr
\end{equation}
\begin{equation}\label{eqn15}
    D_{y,ij}^{(\zeta)}= ie \,\Omega_y \int_{0}^{R} [\chi_{1,i}^{(\zeta) *}(r)\chi_{1,j}^{(\zeta)}(r)+ \chi_{2,i}^{(\zeta) *}(r)\chi_{2,j}^{(\zeta)}(r)] r^2 \,dr
\end{equation}
where $\Omega_{x,y} = \pi [\delta(m_{i},m_{j}-1) \pm \delta(m_{i},m_{j}+1)]$. Here the plus and minus signs correspond to the $x$ and $y$ components of the dipole moments, respectively. Thus, the dipole transitions between states $i$ and $j$ have the following selection rules $m_{i}=m_{j}\pm1$, where 
$m_{i}$ and $m_{j}$ are the corresponding orbital angular momentum quantum numbers [[See Appendix (\ref{app1pt2})]].

The dynamics of a TMDC quantum dot in the field of an ultrafast optical pulse is determined by the following Hamiltonian 
\begin{equation}\label{eqn16}
    \mathcal{H}^{(\zeta)}=\mathcal{H}_{0}^{(\zeta)} + e\textbf{F}(t)\cdot\textbf{r}.
\end{equation}
%%%%%%%%%%%%%%%%%%%%%%%%%%%%%%%%%%%%%%%
\begin{figure}[t!]
    \centering
    \includegraphics[width=8.6 cm]{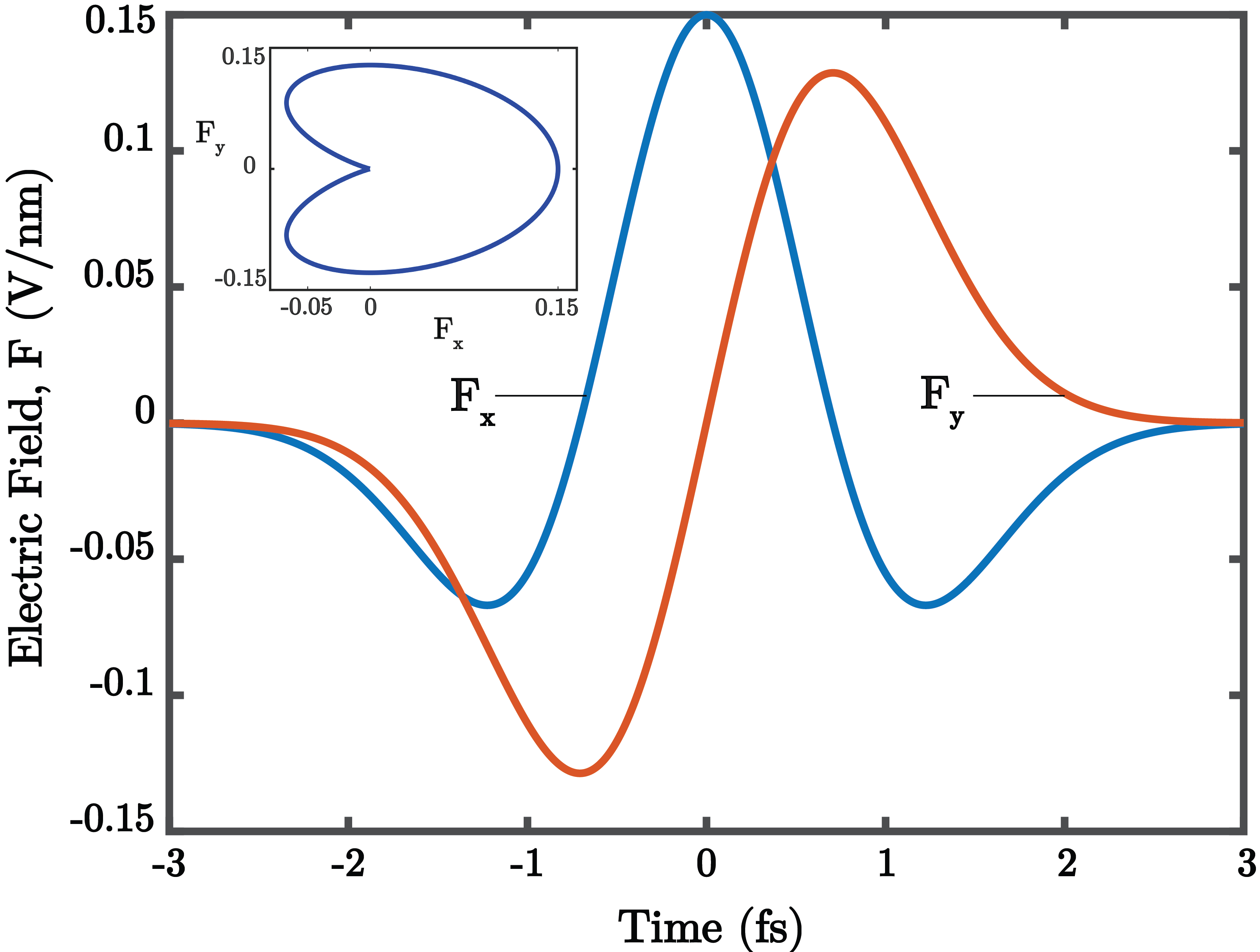}
    \caption{Profile of the electric field of an ultrashort left circularly polarized optical pulse. The pulse consists of a single oscillation. The amplitude of the pulse is $F_{0} = 0.15 V/nm$. Both $x$ and $y$ components of the optical field are shown. The corresponding cycle of the optical pulse is shown in the inset.}
    \label{fig3}
\end{figure}
%%%%%%%%%%%%%%%%%%%%%%%%%%%%%%%%%%%%%%%%%%
Here, $\textbf{F}(t)$ is the electric field of the pulse. 
Below, we consider a circularly polarized optical pulse, which breaks the time-reversal symmetry and thus can generate finite valley polarization.  The $x$ and $y$-components of the 
electric field of a single-oscillation circularly polarized pulse are given by the following expression
\begin{equation}\label{eqn17}
\textbf{F}(t)=
\begin{cases}
        F_{x}(t)=F_{0} \, e^{-u^{2}}(1-2u^{2})\\
         F_{y}(t)=F_{0} \, e^{-u^{2}}(2u)
\end{cases}
\end{equation}
where $u=t/\tau$ and the time parameter $\tau$ determines the frequency and duration of the pulse. The results in the following section are obtained for $\tau=1 fs$, corresponding to a frequency of the pulse of $\approx 1eV$. We consider a single oscillation of the field, and demonstrate the ultrafast control of the valley degree of freedom in TMDC QD system. 
Further, we assume that the characteristic relaxation time for our system are greater than the pulse duration, which in our case is about 6 fs. Thus, the electron dynamics within the TMDC QD is described by the time-dependent Schr\"{o}dinger equation,
\begin{equation}\label{eqn18}
    i\hbar\frac{d\mathbf{\Phi}^{\zeta}(t)}{dt}=\mathcal{H}^{(\zeta)}(t) \mathbf{\Phi}^{(\zeta)}(t).
\end{equation}
We expand a solution of Eq.(\ref{eqn18}) in the basis of 
single-particle states (\ref{eqn3}), $\mathbf{\Psi_{i}^{\zeta}}$,
\begin{equation}\label{eqn19}
    \mathbf{\Phi}^{\zeta}(t)= \sum_{i=1}^{N} \beta_{i}^{(\zeta)}(t) \,\mathbf{\Psi_{i}^{\zeta}}e^{-\frac{i}{\hbar}\varepsilon_{i}^{(\zeta)}t}
\end{equation}
where $\beta_{i}^{(\zeta)}(t)$ are the time-dependent expansion coefficients. Within the low-energy effective model, the energy spectrum of a TMDC QD consists of an infinite number of conduction and valence energy levels. We consider a finite number $N$ of QD energy states, which are in the energy interval from $-1.2 eV$ to $-1.2 eV$. The number of the corresponding QD energy levels depends on the size of the dot, i.e., its radius $R$.
The optical pulses considered below have intensities small enough to prevent inter-valley mixing of QD states. The inter-valley distance in the reciprocal space is $4\pi/3\sqrt{3}a$. 
The electric field, which can transfer an electron over such a distance and result in inter-valley mixing, can be estimated as   $F_{0}\approx 4\pi \hbar/3\sqrt{3}a \approx 0.5$ V/nm. Below  we consider optical pulses with the amplitudes much 
smaller than $0.1$ V/nm. In this case, we can disregard the inter-valley mixing induced by an optical pulse. 
The expansion coefficients in Eq.(\ref{eqn19}), $\beta_{i}^{(\zeta)}(t)$, satisfy the following system of differential equations
\begin{equation}\label{eqn20}
    \frac{d\beta_{m}^{(\zeta)}}{dt}=-\frac{i}{\hbar}\sum_{n=1}^{N} \textbf{F}(t)\cdot\textbf{D}_{mn}^{(\zeta)}(t)\beta_{n}^{(\zeta)}e^{-\frac{i}{h}(\varepsilon_{n}^{(\zeta)}-\varepsilon_{m}^{(\zeta)})t}.
\end{equation}
The system of equations (\ref{eqn20}) was solved numerically under the initial conditions that all the conduction band (CB) states are empty and all the valence band (VB) states are occupied. The numerical solutions were obtained using the explicit Runge-Kutta method with the initial time being -3$\tau$ and the final time being 3$\tau$. With the known solution of Eq.(\ref{eqn20}) we can find the populations of the CB states for both $K$ and $K'$ valleys,
\begin{equation}\label{eqn21}
    \mathcal{N}^{(\zeta)}_{CB}(t)=\sum_{j \in CB}|\beta_{j}^{(\zeta)}(t)|^{2}.
\end{equation}
With the known CB populations of different valleys, the valley polarization is defined through the following expression 
\begin{equation}\label{eqn22}
    P_{N}=\frac{\mathcal{N}_{CB}^{(K)}(t=\infty)-\mathcal{N}_{CB}^{(K')}(t=\infty)}{\mathcal{N}_{CB}^{(K)}(t=\infty)+\mathcal{N}_{CB}^{(K')}(t=\infty)}.
\end{equation}
Here, when electrons are excited in one valley only, the value of $P_{N}$ has a maximum value of 1.

We also define the average density of states, $\mathcal{G}^{\zeta}$, which quantifies, for a given valley ($K$ or $K'$), the density of energy eigenstates in the conduction or valence band with a given spin configuration. Such average density of states is defined through the following expression 
\begin{equation}\label{eqn23}
    \mathcal{G}^{\zeta}= \frac{\mathcal{D}_N^{\zeta}}{\Xi^{\zeta}}
\end{equation}
where $\mathcal{D}_N^{\zeta}$ is the number of spin-dependent energy eigenstates in a given band, and $\Xi^{\zeta}$ is the difference between the maximum and the minimum energy values in the band.

%%%%%%%%%%%%%%%%%%%%%%%%%%%%%%%%%%%%%%%%%%%%%%
%MODEL AND MAIN EQUATIONS SECTION ENDS
%%%%%%%%%%%%%%%%%%%%%%%%%%%%%%%%%%%%%%%%%%%%
\section{RESULTS AND DISCUSSION}\label{sec3}
\begin{figure}[t!]
    \centering
    \includegraphics[width=8.6 cm]{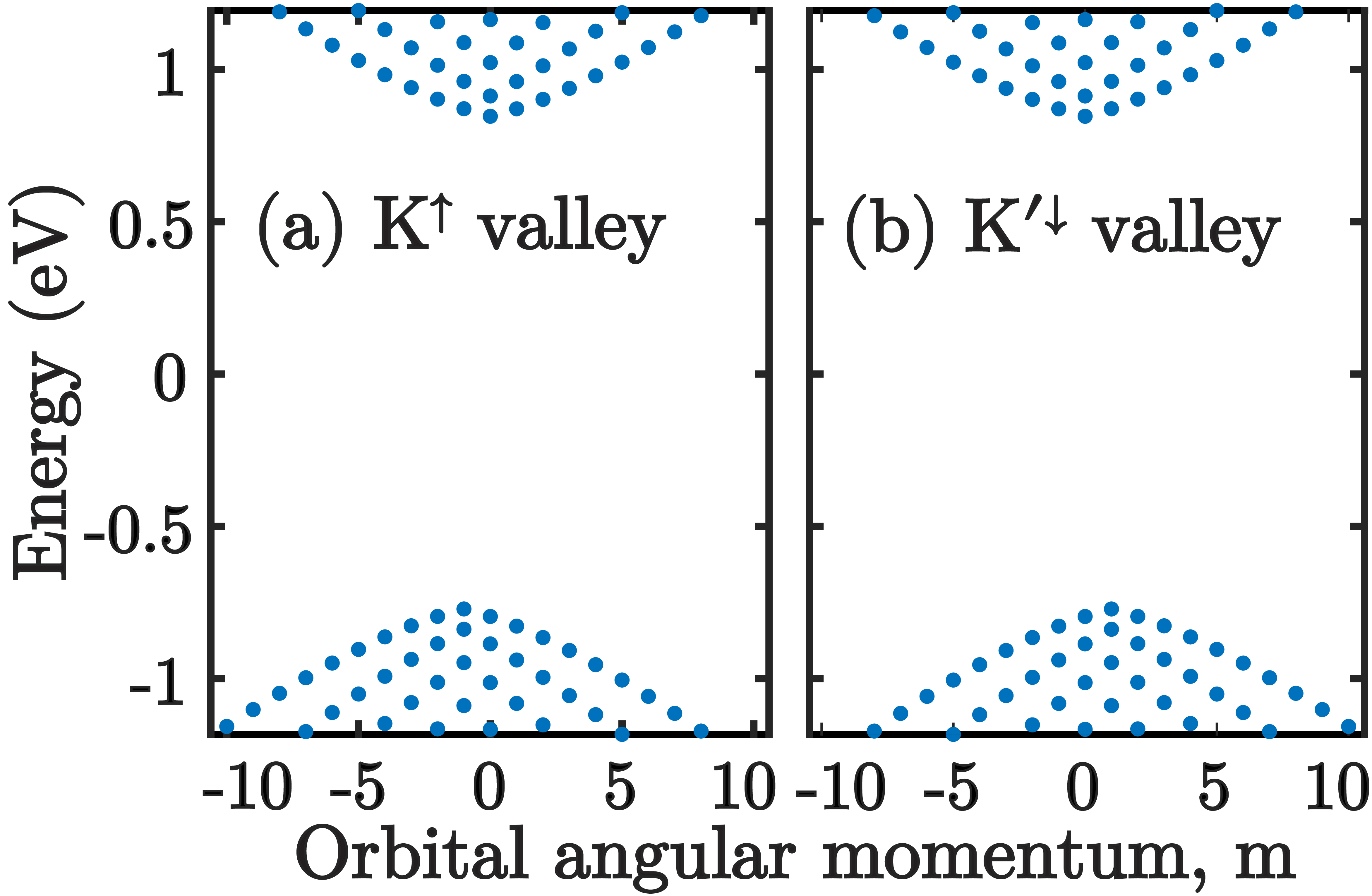}
    \caption{Energy spectra of $K^{\uparrow}$ valley (a) and $K'^{\downarrow}$ valley
    (b) of MoS\textsubscript{2} QD with the radius of $R$=5 nm. For both valleys, the spectrum is shown as a function of the orbital angular momentum, $m$, which takes integer values.}
    \label{fig4}
\end{figure}

\begin{figure}[t!]
    \centering
    \includegraphics[width=8.6 cm]{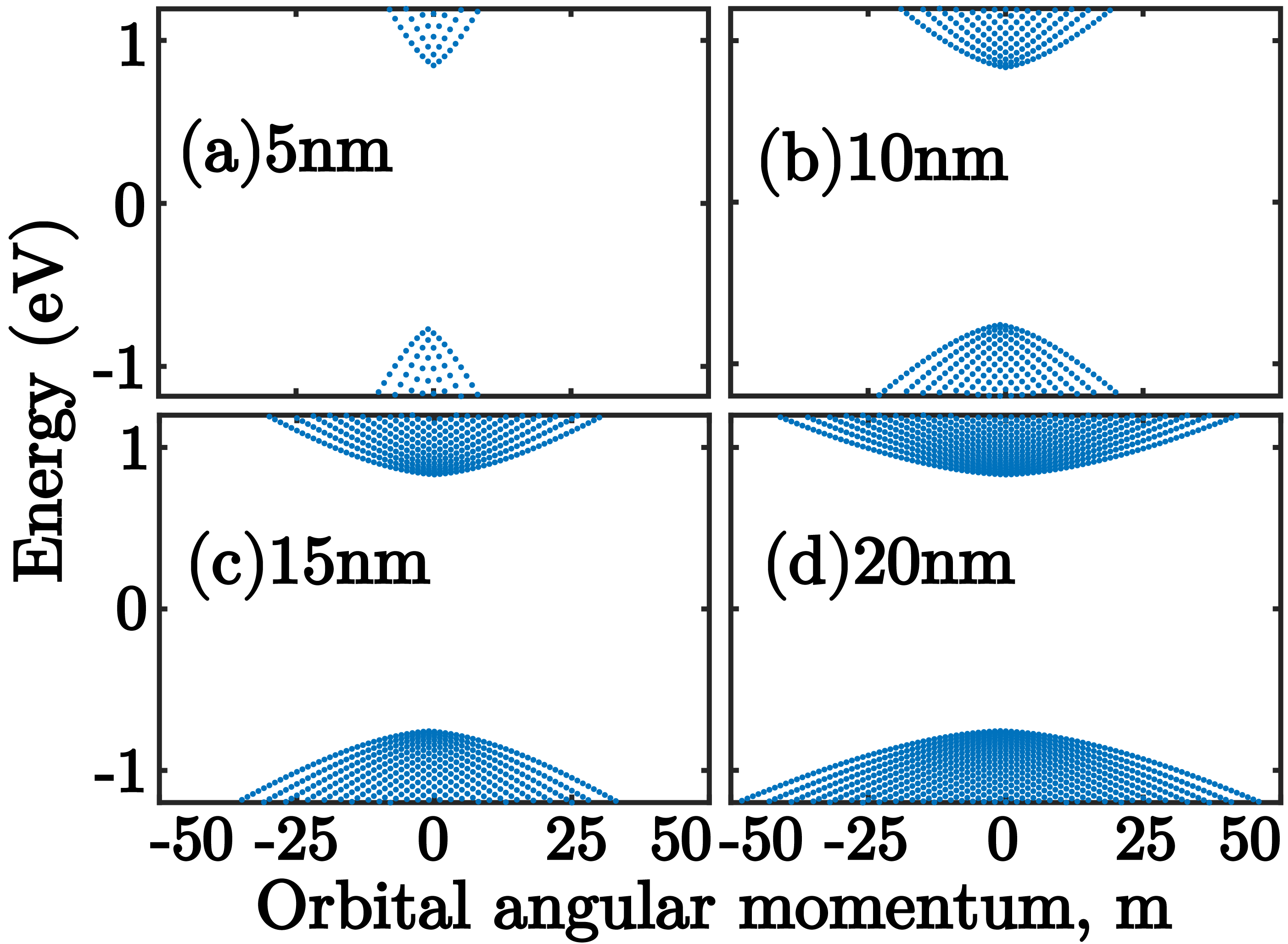}
    \caption{Energy spectra of MoS\textsubscript{2} QD ($K^{\uparrow}$ valley) with the radius of $R$= (a) $5$ nm, (b) $10$ nm, (c) $15$ nm, and (d) $20$ nm. For all cases, the spectrum is shown as a function of the orbital angular momentum, $m$, which takes integer values.}
    \label{fig5}
\end{figure}

The energy spectra of a MoS\textsubscript{2} QD, calculated using Eq.(\ref{eqn12}), are shown for $K^{\uparrow}$ and $K'^{\downarrow}$ valleys in Fig.(\ref{fig4}). The spectra demonstrate that the condition of the time-reversal symmetry is satisfied for the valleys with opposite spin states, i.e., $ \varepsilon^{\uparrow}_{K,m}= \varepsilon^{\downarrow}_{K',-m}$. The QD spectra have finite bandgaps, which depend on the size of the system.  
The states with the positive and negative energies 
correspond to the conduction band and the valence band states, respectively. The energy spectra of MoS\textsubscript{2} quantum dots with different radii are shown in Fig.(\ref{fig5}) for the 
$K^{\uparrow}$ valley. 
As expected, with increasing the QD radius, the number of energy levels within the fixed energy window increases and the band gap decreases.

Another important feature of TMDC QD energy spectra is that the maxima of the valence band energy spectra occur at nonzero values of $m$, and the corresponding values of $m$ have opposite signs at the $K$ and $K'$ valleys. Such a property results in different responses of two valleys to a circular polarized pulse with a specific helicity. 
For example, if an incident pulse is clockwise circularly polarized, then the dipole selection rule is 
$m_{final}-m_{initial}=1$. The energy difference, $\Delta\varepsilon =\varepsilon_{final} -\varepsilon_{initial}$ determines the rate of the corresponding optical transitions.
Since $\Delta\varepsilon$ is different for the $K$ and $K'$ valleys due to the different behaviors of the energy spectra of the valleys, the dipole couplings for the two valleys will different, which finally results to different generated conduction band populations.

\begin{figure}[b]
    \centering
    \includegraphics[width=8.6 cm]{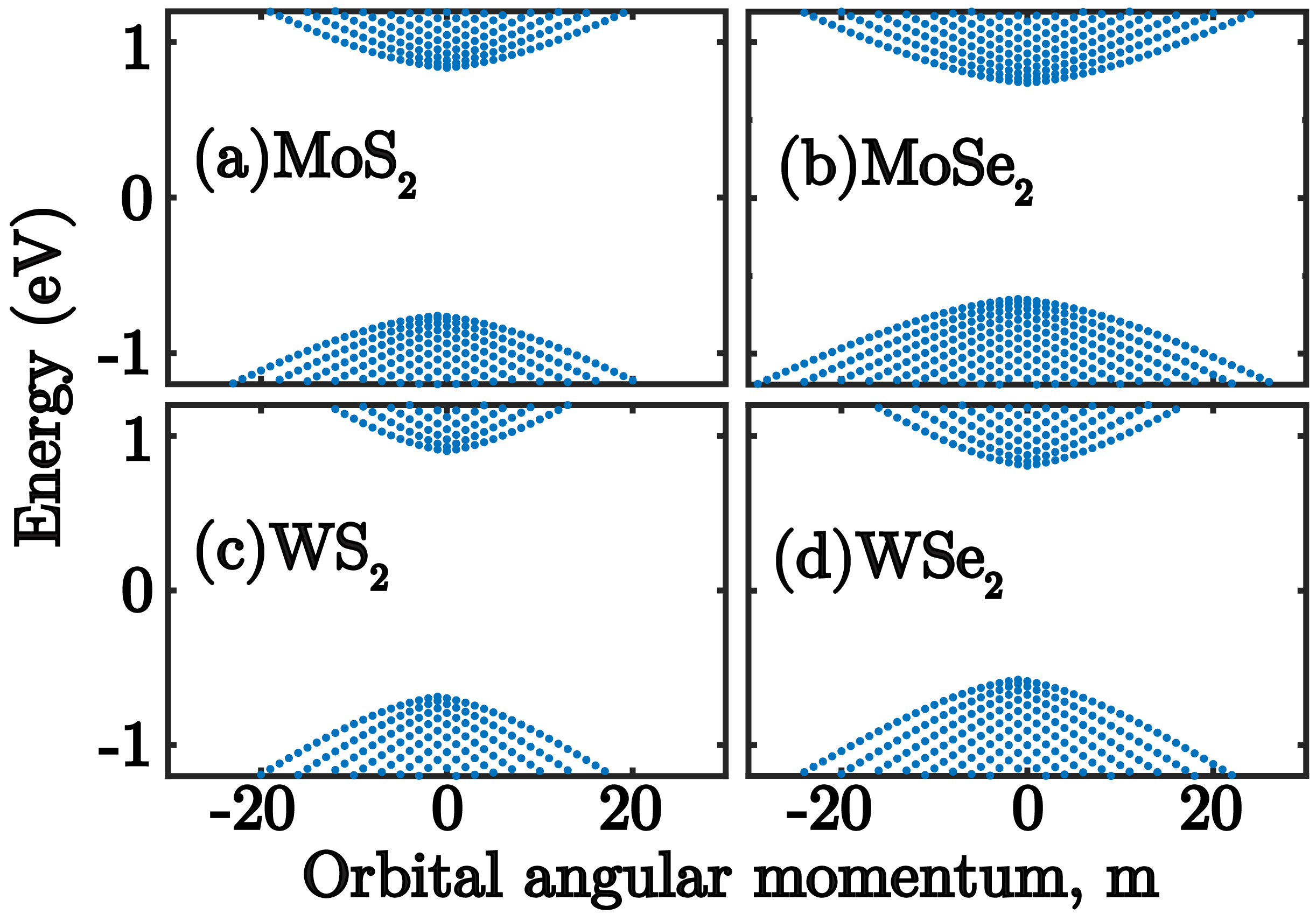}
    \caption{Energy spectra of different TMDC QDs - (a) MoS\textsubscript{2} QD, (b) MoSe\textsubscript{2} QD, (c) WS\textsubscript{2} QD, and  (d) WSe\textsubscript{2} QD. For all materials, the spectrum is shown as a function of the orbital angular momentum $m$. The QD radius is $R$=10 nm.}
    \label{fig6}
\end{figure}

Figure (\ref{fig6}) shows the energy spectra for different TMDC materials: (a) MoS\textsubscript{2} (b) MoSe\textsubscript{2} (c) WS\textsubscript{2} (d) WSe\textsubscript{2} for the QD  radius of $\textbf{R}=10$ nm. 
Among all TMDC materials, WS\textsubscript{2} has the lowest number of the total energy eigenstates within the considered energy range, as well as the smallest range of allowed $m$ values. It also has the smallest number of available CB energy states. For other three TMDC materials, the energy spectra have the same general structure and satisfy the spin-dependent condition for time-reversal symmetry. Further, between themselves, the quantum dots for materials that have the chalcogen atom as Se, namely MoSe\textsubscript{2} and WSe\textsubscript{2}, have smaller band gaps compared to their counterparts with S as the chalcogen atom. This is also consistent with the monolayer band gap values, as given in Table \ref{tab1}.

\begin{figure}[t!]
    \centering
    \includegraphics[width=8.6 cm]{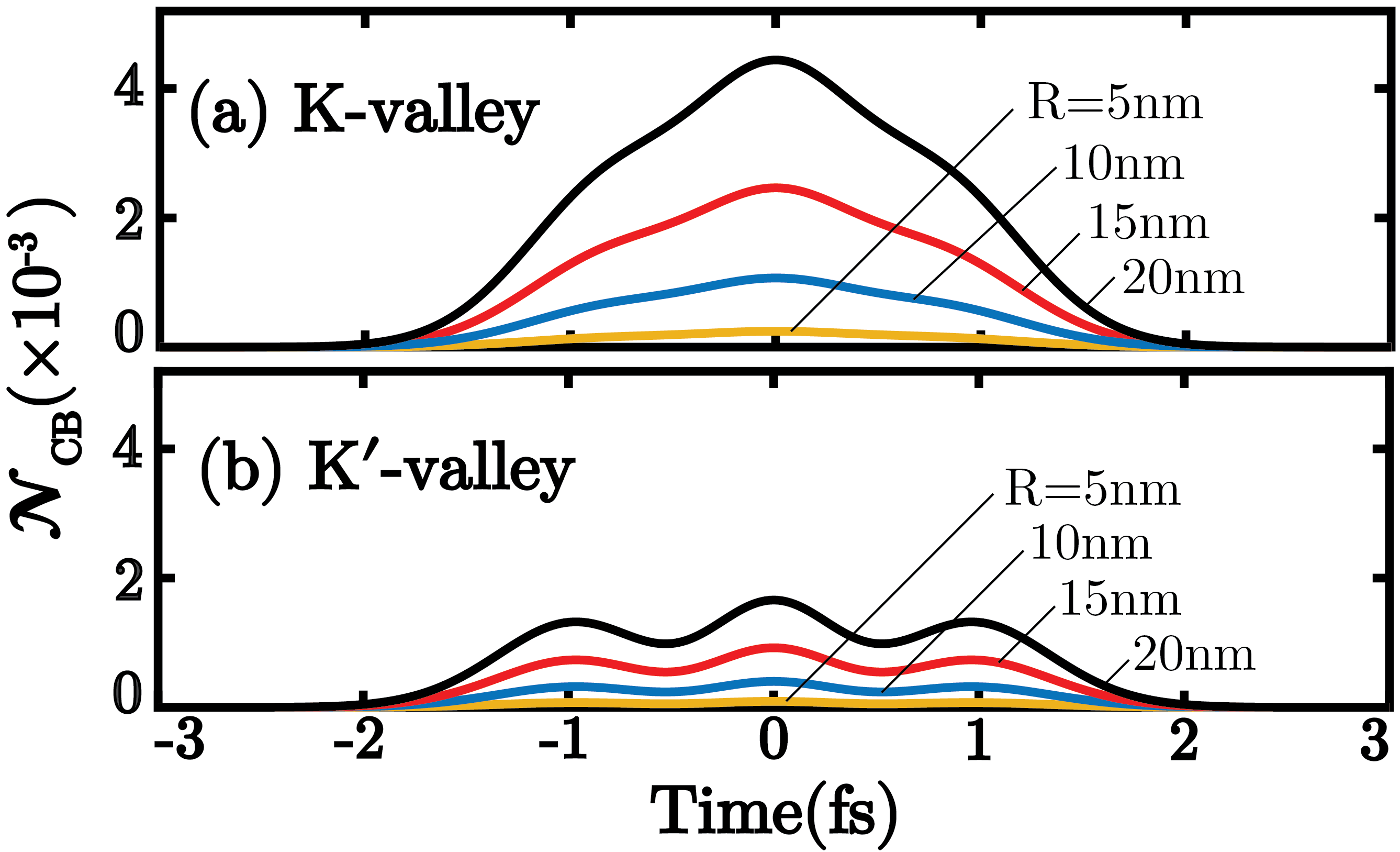}
    \caption{Conduction band population of MoS\textsubscript{2} QDs as a function of time. The positive energy states of QDs correspond to the conduction band. The data are shown for different QD radii for (a) the $K$ valley (b) the $K'$ valley. 
 The corresponding values of QD radius are shown next to each line.  The pulse considered is left circularly polarized with the amplitude of $F_{0} =0.05$ V/nm.}
    \label{fig7}
\end{figure}

As noted in the previous section, the effective low-energy model of TMDC QDs generates an infinite number of energy levels. In Figs. (\ref{fig4}), (\ref{fig5}), and (\ref{fig6}), only the levels that are considered below are shown. 
They are within the energy interval of $-1.2$ eV $<\varepsilon<1.2$ eV. Such an energy interval is chosen to 
guarantee the convergence of the results, namely, the total conduction band population for the field amplitudes of up to $F_0=0.15$V/nm.

\begin{figure}[b]
    \centering
    \includegraphics[width=8.6 cm]{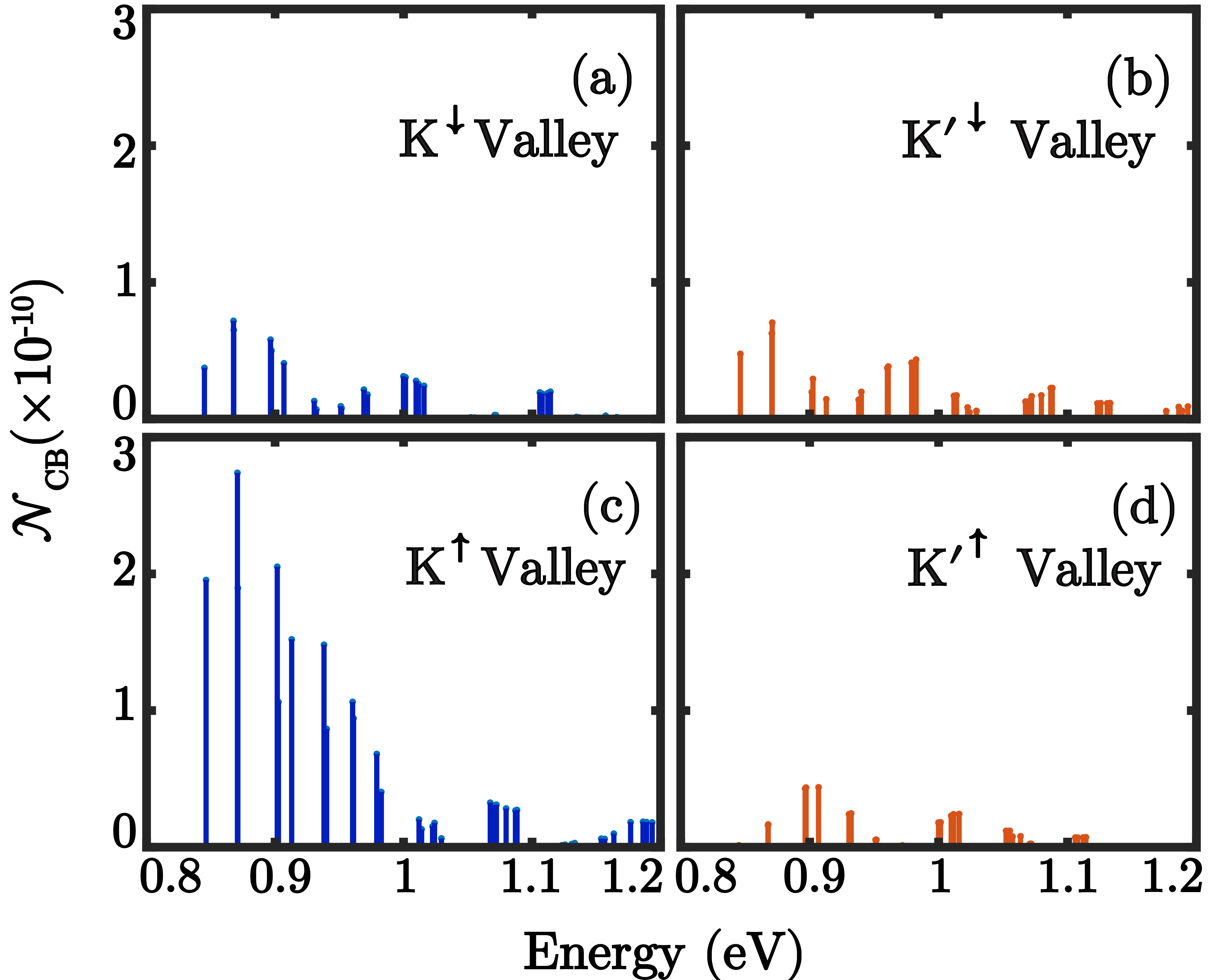}
    \caption{Population of individual conduction band levels for different spin components of  MoS\textsubscript{2} QD of radius $R=5$ nm. The results are shown for (a) $K^{\downarrow}$ valley, (b)  $K'^{\downarrow}$ valley, (c)  $K^{\uparrow}$ valley, and (d) $K'^{\uparrow}$ valley. The amplitude of the pulse is $F_{0}= 0.05$ V/nm. 
The QD radius is $R=5$ nm. The optical pulse is left circularly polarized.}
    \label{fig8}
\end{figure}

To study the induced ultrafast valley polarization of TMDC QDs, we apply a single cycle of a left-circularly polarized optical pulse. The $x$ and $y$ components of the pulse, i.e., the pulse profile, are shown in Fig. \ref{fig3}. Before the pulse, all the valence band states, i.e., the states with negative energies, are populated. One of the  characteristics of the electron dynamics in the field of an optical pulse is the total CB population, $\mathcal{N}_{CB}^{\zeta}$. 
In Fig. \ref{fig7}, the total CB populations for MoS\textsubscript{2} QDs are shown a function of time for two valleys, $K$ and $K'$, and for different QD sizes, $R=$ 5 nm, 10 nm, 15 nm, and 20 nm. 
The field amplitude $F_0$ of the pulse is $0.05$V/nm. For all QDs, the electron dynamics is highly reversible, i.e., the residual CB population is much less than the maximum CB population during the pulse and is almost the same as the initial CB population. Similar dynamics of the CB population has been observed in TMDC monolayers\cite{RN227}. 
Although the residual CB population is small, it is different for two valleys, which finally results in the final valley polarization of TMDC QDs. 
For all systems, the residual CB population 
for the $K$ valley is higher than 
the one for the $K'$ valley for all values of the radius, which results in finite valley polarization of TMDC QDs.

\begin{figure}[t]
    \centering
    \includegraphics[width=8.6 cm]{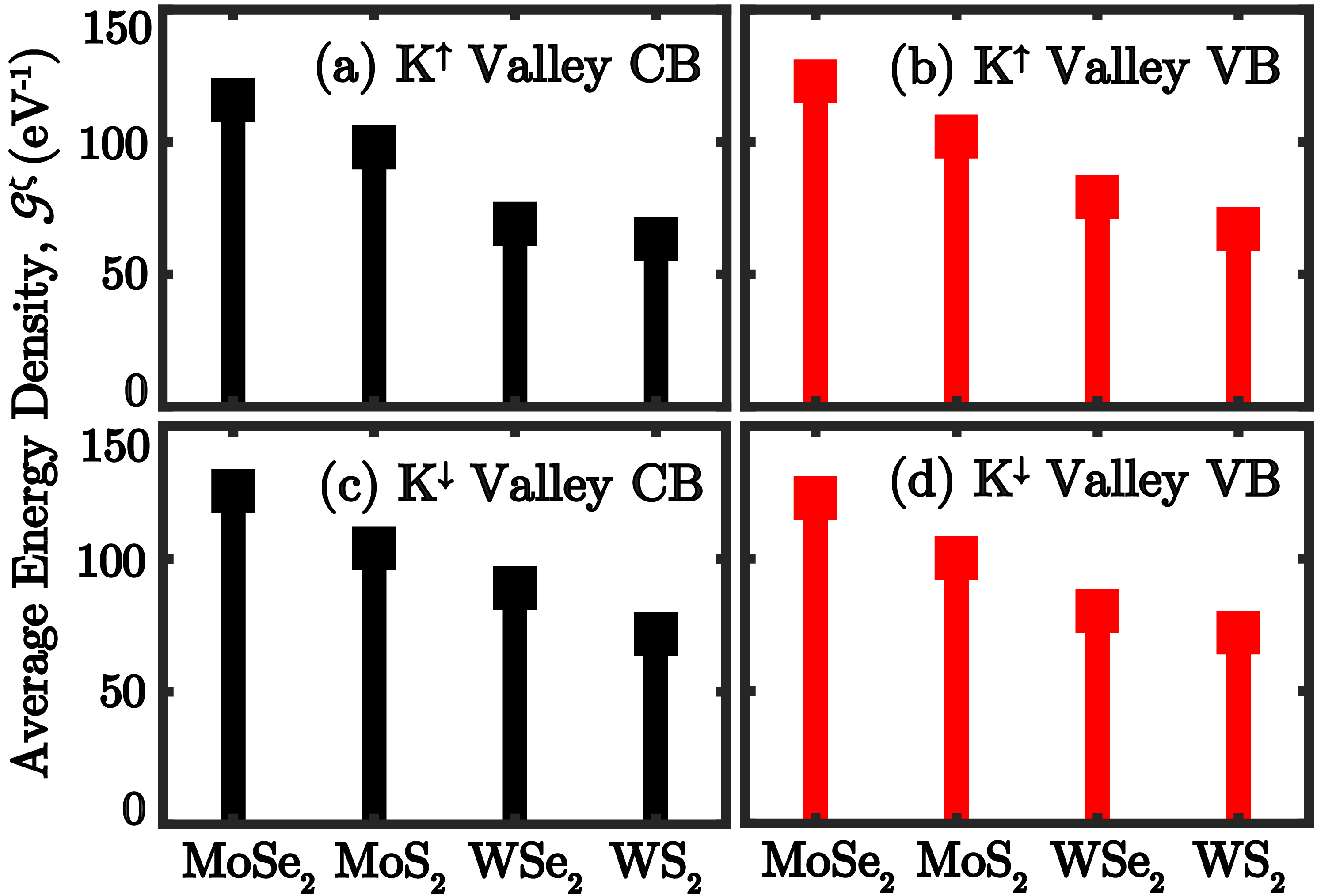}
    \caption{Average density of states for different spin components and different TMDC materials. The density of states is shown for (a) $K^{\uparrow}$ valley of the conduction band; (b) $K^{\uparrow}$  valley of the valence band; (c) $K^{\downarrow}$ valley of the conduction band; and  (d) $K^{\downarrow}$ valley of the valence band. The QD radius is   $R=5$ nm. }
    \label{fig9}
\end{figure}

The residual populations of individual CB levels are shown for a MoS\textsubscript{2} quantum dot of radius $\textbf{R}=$ 5 nm in Fig. \ref{fig8}. The populations characterize the residual state of the system. The results are shown for two valleys, $K$ and $K'$, and for two spin components, spin-up and spin-down. The field amplitude of the applied pulse is $F_0 = 0.05$ V/nm. As expected, the CB states with the low energies are more populated than the high energy states. This effect is most pronounced for the $K^{\uparrow}$ valley. Also, while the populations of the states of the $K'$ valley are almost the same for both spin components, for the $K$ valley, mainly the spin-up component is populated. 
The residual CB populations clearly show the finite valley polarization, where the $K$ valley is more populated than the $K'$ valley. The results shown in Fig.\ \ref{fig8} are typical for TMDC QDs. 

%\begin{figure}[t]
%    \centering
%    \includegraphics[width=8.6 cm]{IllFileProcAvgEnergyDensityAllTMDCRadCompareCBSpinup.png}
%    \caption{Average energy density of conduction band single-particle energy states of different transition metal dichalcogenide materials at radii \textbf{R} $=$ 5,10,15,20 nm. The results show energy densities for (a) MoS\textsubscript{2} (b) MoSe\textsubscript{2} (c) WS\textsubscript{2} (d) WSe\textsubscript{2}. All results shown are for $K^{\uparrow}$ - valley conduction band.}
%    \label{fig10}
%\end{figure}

One of the characteristics of the energy spectra of TMDC QDs is the average energy density of the conduction and valence bands. Such average energy density also determine the response of the system to external optical field. 
In Fig. (\ref{fig9}), the average energy densities calculated for spin-up and spin-down components are shown for the conduction and valence bands. The QD radius is 5nm. The average energy density $\mathcal{G}^{\zeta}$ is calculated as the ratio of the number of the energy states in a given band and the difference of the maximum and minimum energy values in the corresponding band. The data show that the average energy density monotonically decreases following the order MoSe\textsubscript{2} $>$ MoS\textsubscript{2}$>$ WS\textsubscript{2}$>$ WSe\textsubscript{2}. This is consistent with the parameters of TMDC materials shown in Table \ref{tab1}. This trend is similar for different spin configurations and different bands.

The main outcome of interaction of a circularly polarized optical pulse with TMDC QDs is the residual valley polarization of the system. The valley polarization  $P_N$ defined by  Eq.~(\ref{eqn22}) is shown for QDs of different TMDC  materials: (a) MoS\textsubscript{2}, (b) MoSe\textsubscript{2}, (c) WS\textsubscript{2}, and  (d) WSe\textsubscript{2}. 
At small field amplitudes, the valley polarization approximately increases as the third power of $F_0$. 
The valley polarization shows saturated behavior at larger values of $F_0$, which is due to the finite number of energy levels in the considered QD systems. In general, at small field amplitudes, with increasing the QD size, the valley polarization increases. Under such conditions, to have large valley polarization, the QD should have large size. Different behavior is realized for large field amplitude, $F_0>0.05$ V/nm. In this case, the dependence of the valley polarization on the QD radius is non-monotonic and, for some cases, the large valley polarization is realized for intermediate QD sizes. For example, for MoS$_2$ QDs and the field amplitude of 0.1 V/nm, the largest valley polarization is realized for the QD with the radius of 15 nm.

\begin{figure}[t!]
    \centering
    \includegraphics[width=8.6 cm]{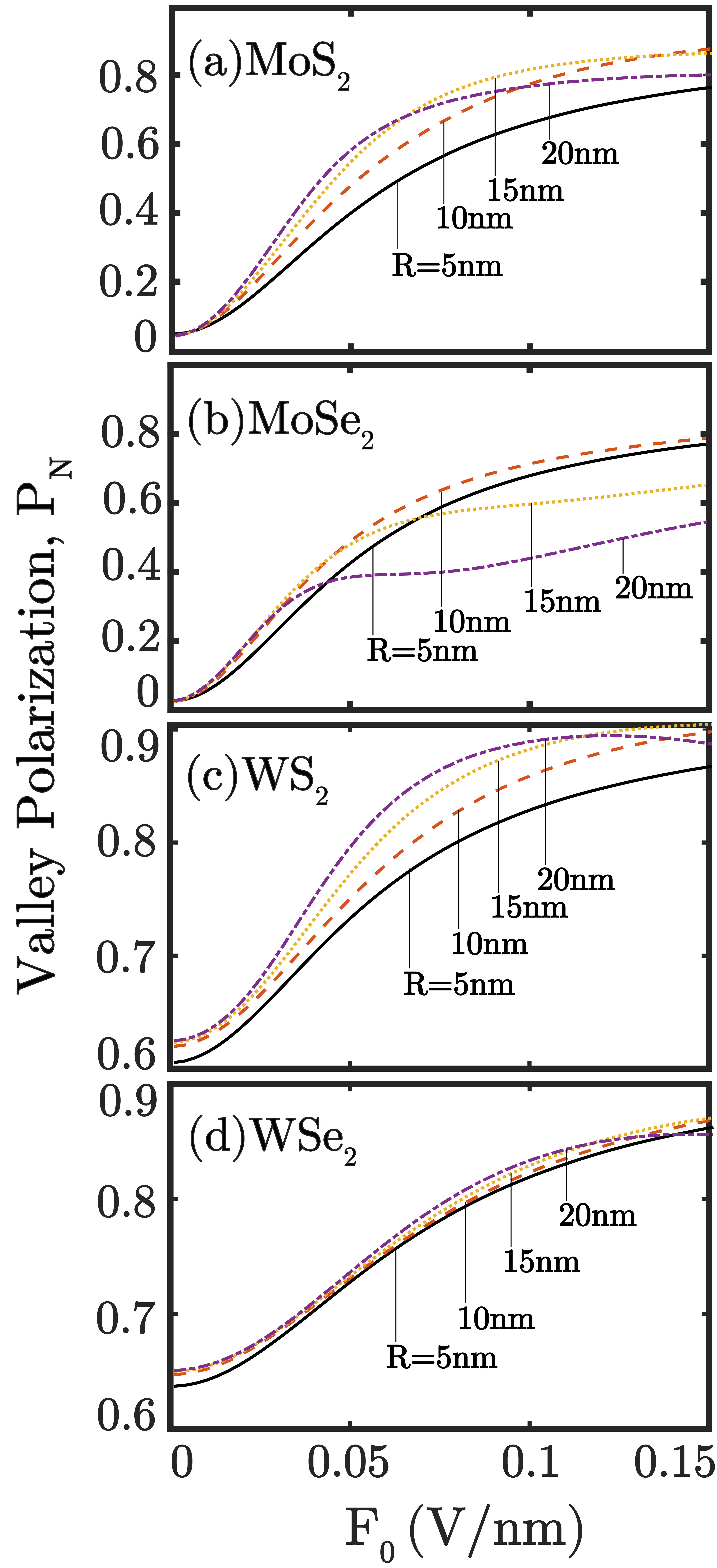}
    \caption{The valley polarization P\textsubscript{N} of TMDC QDs. 
    The results are shown for (a) MoS\textsubscript{2}, (b) MoSe\textsubscript{2}, (c) WS\textsubscript{2}, and (d) WSe\textsubscript{2} QDs. The QD radii are shown by numbers next to the corresponding lines. The optical pulse is left-circularly polarized. }
    \label{fig12}
\end{figure}

\begin{figure}[t!]
    \centering
    \includegraphics[width=8.6 cm]{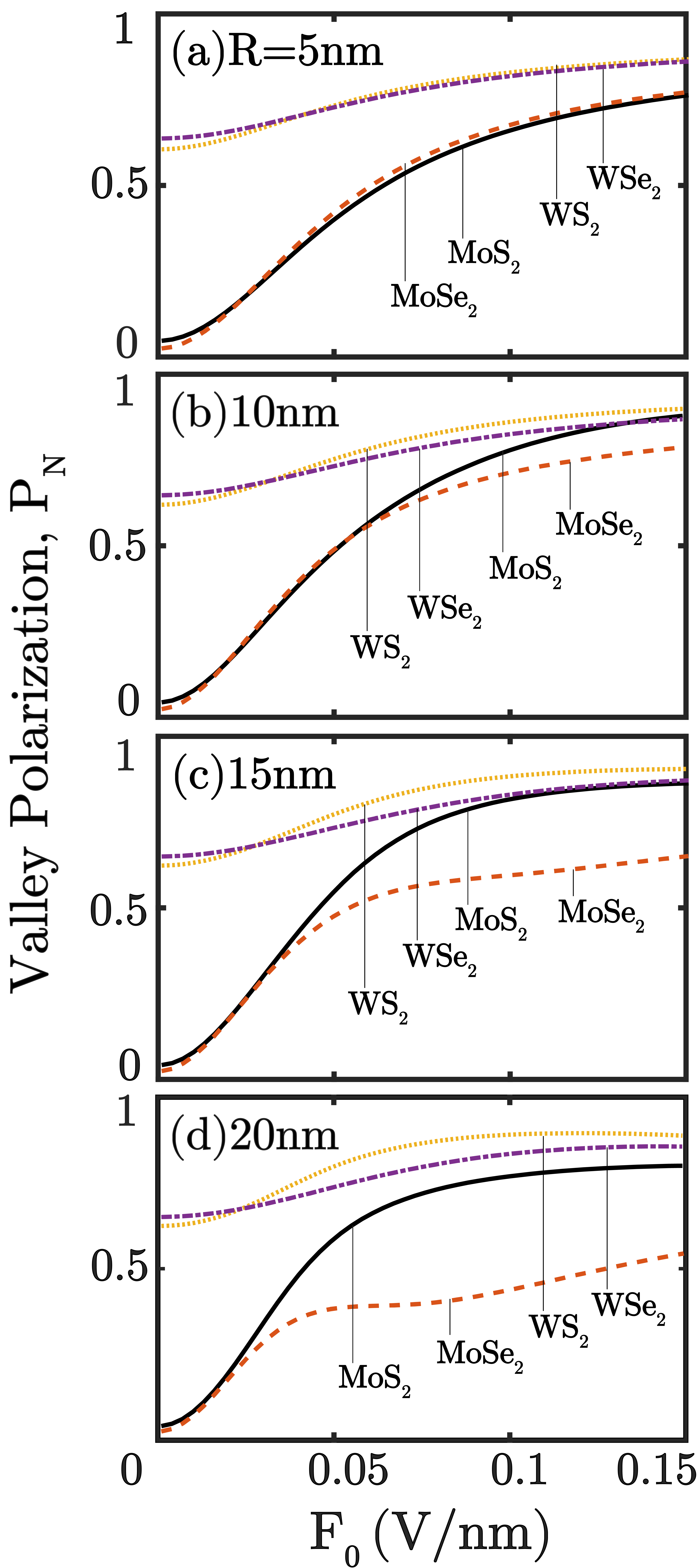}
    \caption{The valley polarization, P\textsubscript{N}, as a function of the field amplitude. The results are shown for QD radius of (a) $R=5$ nm; (b) $R=10$ nm; (c) $R=15$ nm, and (d) $R=20$ nm. For each radius, the TMDC materials are marked next to the corresponding lines. The optical pulse is left-circularly polarized.}
    \label{fig13}
\end{figure}

To compare the valley polarizations of different TMDC materials, we plot in Fig.\ref{fig13} the valley polarization as a function of the field amplitude at a fixed  QD radius but for different TMDC materials. For all QD sizes, for small field amplitudes, $F_0\ll 0.05$ V/nm, the  valley polarization for WS\textsubscript{2} and WSe\textsubscript{2} is at around $60$ \% and saturates at values greater than  $90$ \% at the higher field amplitudes. At the same time, the valley polarization for MoS\textsubscript{2} QDs is less than $10$ \% at $F_0 \rightarrow 0$ and reaches its saturated value at around $60-80$ \%. Also, for large QD sizes, i.e., for $R=15$ nm and 20 nm, the saturated valley polarization is much smaller than the one for other TMDC materials. This property is related to the fact that  MoSe\textsubscript{2} possesses the highest average density of states, $\mathcal{G}^{\zeta}$.
This leads to a larger number of energy eigenstates of the MoSe\textsubscript{2} quantum dot being involved in the interband transitions during the pulse, compared to other TMDC materials, and this effect becomes more pronounced as the radius and correspondingly   the number of energy eigenstates increases.

\begin{figure}[b]
    \centering
    \includegraphics[width=8.6 cm]{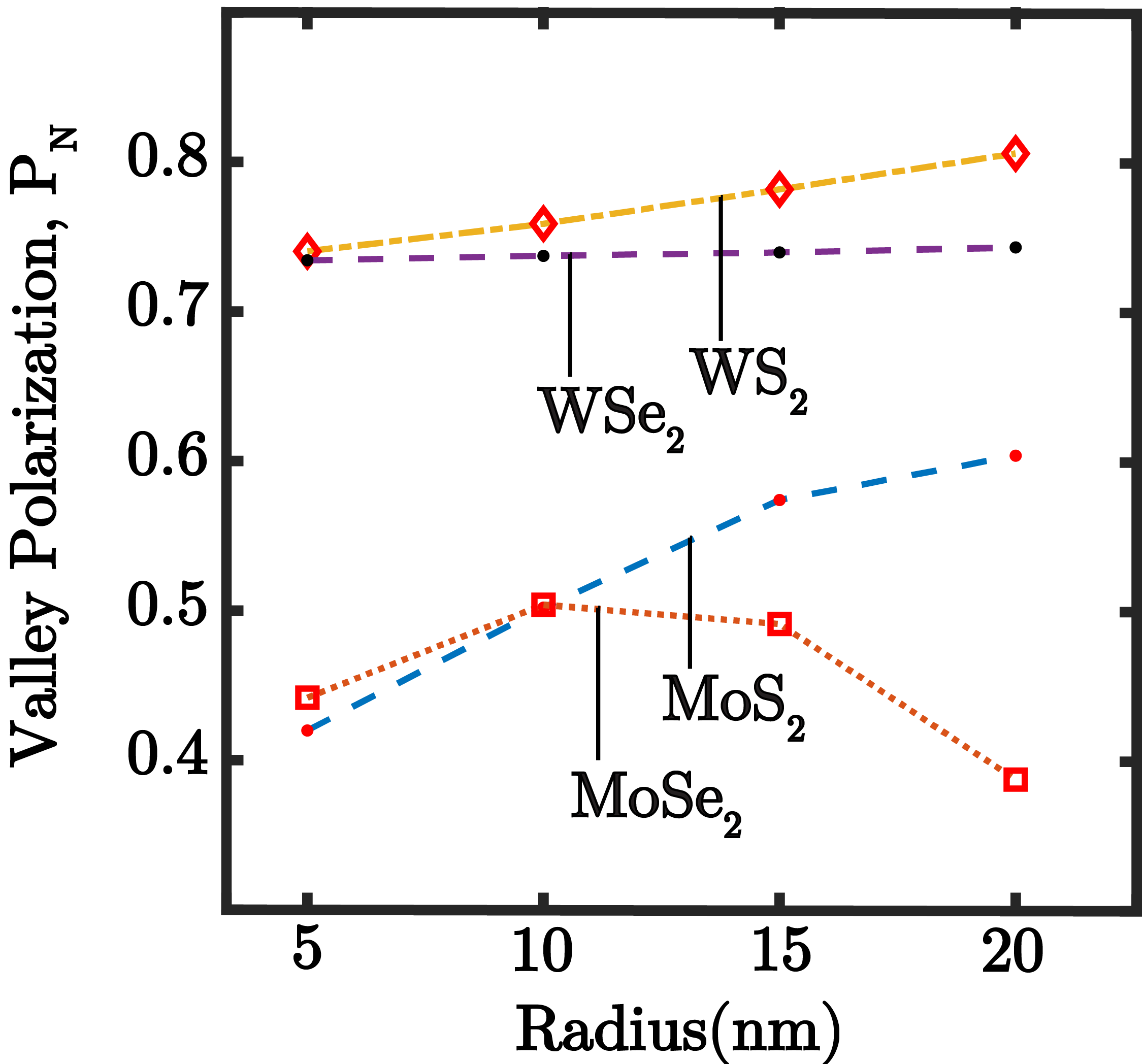}
    \caption{The valley polarization, P\textsubscript{N}, as a function of the QD radius. 
    The pulse amplitude is $F_{0}=0.05$ V/nm.  The TMDC materials are marked next to the corresponding lines. The optical pulse is left-circularly polarized. }
    \label{fig11}
\end{figure}

For a relatively low amplitude of the optical pulse, $F_0 = 0.05$ V/nm, the dependence of the valley polarization on the QD radius is shown in Fig.\ \ref{fig11} for different TMDC materials. The data show that the valley polarization for WSe$_2$ and WS$_2$ is in the range of 70-80 \%, while the valley polarization for MoS$_2$ and MoS$_2$ is in the range of 40-60 \%. The behavior of valley polarization follows two sets of data, WSe$_2$ and WS$_2$ form one set and MoS$_2$ and MoS$_2$ form another set. Such similarity maybe related to the similar values of the average density of states for these materials, see Fig.\ \ref{fig9}, i.e., MoS$_2$ and MoS$_2$ have density of states close to 100-120 eV$^{-1}$, while the density of states of WSe$_2$ and WS$_2$ is in the range of 50-80 eV$^{-1}$.   
 As a function of the QD radius, the valley polarization generally increases with $R$ reaching its maximum value at $R=20$ nm. There is one exception from this rule. Namely, for MoSe$_2$, the valley polarization has maximum at intermediate radius of 10 nm. Also, the strongest dependence of the valley polarization on the radius is realized for MoS$_2$, while the weakest dependence is visible for WSe$_2$, for which the valley polarization is almost independent on the QD size. Thus, the dependence of the valley polarization on the QD radius is strongly sensitive to the material of QD. Depending on the TMDC material, the valley polarization shows very weak dependence on QD radius or strong monotonic increase with $R$ or non-monotonic dependence with maximum valley polarization realized at finite value of $R$.

\section{CONCLUSION}\label{sec4}

In transition metal dichalcogenide materials with two inequivalent valleys, the valley degree of freedom can be controlled by an ultrashort optical circularly polarized pulse, which brakes the time reversal symmetry of the system. The valley polarization induced by such a pulse can be also tuned by introducing spatial confinement of the carries, i.e., by considering QDs of TMDC materials. The size of a QD control the bangap of the system and effective average density of states, which finally affects how an external optical pulse interacts with the material. For QDs considered in the present work, the controllable parameter is the QD radius, $R$. The dependence of the valley polarization on $R$ is not universal and strongly depends on the TMDC materials. For some TMDC materials, the valley polarization is almost insensitive to the QD size, while for other materials, the valley polarization shows strong dependence on $R$, which can be either monotonic increase with $R$ or non-monotonic dependence with local maximum at a finite radius. 

The valley polarization in TMDC QDs can be generated by an optical pulse with a relatively small field amplitude ($F_0 \leq 0.15$ V/nm), and, as expected, within this amplitude range, monotonically increases with $F_0$ reaching a saturated value at large pulse intensity, $F_0 \sim 0.05$ V/nm.  
Among the TMDC materials considered here, very high valley polarization ($~90\%$) can be induced in WS\textsubscript{2} and WSe\textsubscript{2}. The fact that such high valley polarization can be induced by relatively low-intensity pulses and the ability to control the valley polarization by changing the QD radius, provides an effective way to control the valley degree of freedom in transition metal dichalcogenide quantum dot systems.

We have considered here a particular shape of TMDC QDs. We expect that by using a circularly polarized ultrashort optical pulse, one can control the valley polarization in any shape of a finite TMDC system.  Our analysis also does not consider the phenomenon of inter-valley mixing, which can be important for QDs of relatively small sizes, less than 5 nm.

\section{ACKNOWLEDGEMENTS}
Major funding was provided by Grant No. DE-FG02-01ER15213
from the Chemical Sciences, Biosciences and Geosciences
Division, Office of Basic Energy Sciences, Office of Science,
US Department of Energy.
Numerical simulations were performed using support by Grant No. DE-SC0007043
from the Materials Sciences and Engineering Division of
the Office of the Basic Energy Sciences, Office of Science,
US Department of Energy.
%\blindtext \cite{article-minimal}

\section{APPENDIX}\label{app1}
\subsection{Valley dependent conditions on orbital angular momentum, $m$}\label{app1pt1}

From Eqs. (\ref{eqn6}) and (\ref{eqn8}), we can find the explicit  expressions for $\chi_1^{(\zeta)}(r)$. Such expressions depend on the sign of the quantity $(m+1)$, where $m$ is the orbital angular momentum quantum number. 

\subsubsection{Case I: $m+1>0$}

Under this condition, for either value of $\zeta$, we have the following expression 
\begin{equation}
\begin{split}\label{eqn24}
    \bigl(\mathbf{\nabla}_r + \displaystyle\frac{(m+1)}{r}\bigr) J_{m+1}(\xi_{\zeta}r)\\
    =\mathbf{\nabla}_r J_{m+1}(\xi_{\zeta}r) + \frac{(m+1)}{r} J_{m+1}(\xi_{\zeta}r)\\
    = \xi_\zeta J_{m+1}(\xi_{\zeta}r),
\end{split}
\end{equation}
where we have used the well-known Bessel function identity
\begin{equation}\label{eqn25}
    J_n(x)=\pm J^{'}_{n \pm 1}(x) + \frac{(n \pm 1)}{x}J_{n \pm 1}(x).
\end{equation}
Here  $J^{'}_{n \pm 1}(x) = \frac{dJ}{dx}$.

\subsubsection{Case II: $m+1<0$}

Under this condition, we get 
\begin{equation}\label{eqn26}
    \begin{split}
        \bigl(\mathbf{\nabla}_r - \displaystyle\frac{|m+1|}{r}\bigr) J_{|m+1|}(\xi_{\zeta}r)\\
        =\bigl(\mathbf{\nabla}_r  J_{|m+1|}(\xi_{\zeta}r - \displaystyle\frac{|m+1|}{r}\bigr) J_{|m+1|}(\xi_{\zeta}r).
    \end{split}
\end{equation}
Here, $|m+1|+1=-(m+1)+1=-m=|m|$. Then Eq. (\ref{eqn27}) becomes
\begin{equation}\label{eqn27}
    \begin{split}
        \bigl(\mathbf{\nabla}_r - \displaystyle\frac{|m+1|}{r}\bigr) J_{|m+1|}(\xi_{\zeta}r)\\
        =\bigl(\mathbf{\nabla}_r  J_{|m|-1}(\xi_{\zeta}r - \displaystyle\frac{|m|-1}{r}\bigr) J_{|m|-1}(\xi_{\zeta}r)\\
        =-\xi_{\zeta} J_{|m|}(\xi_{\zeta}r),
    \end{split}
\end{equation}
where we have used identity (\ref{eqn25}) to simplify the above expression. 
%To find the expression when $m+\zeta=0$, we note that under this condition $J_{m+\zeta}=J_{0}$ and applying the Bessel function identity gives us, for each valley, the appropriate sign for the expression of $\chi_1^{\zeta}(r)$. Working similarly for $K'$ valley, we can find the corresponding expression for $\chi_1^{\zeta}(r)$ under the relevant conditions on ($m+\zeta$) in that valley.

\subsection{Selection rules on Dipole transitions}\label{app1pt2}

The angular part of the integral in Eq. (\ref{eqn13}) determines the selection rules for dipole transitions between the $i$-th and $j$-th energy eigenstates.
For the matrix elements of the $x$-component of the dipole moment, $ D_{x,ij}^{(\zeta)}$,  the angular part can be isolated as follows
\begin{equation}\label{eqn28}
    \int_{-\pi}^{pi} e^{i(m_j-m_i)\theta} \cos{\theta}  d\theta,
\end{equation}
which becomes 
\begin{equation}\label{eqn29}
\begin{split}
    \frac{1}{2} \int_{-\pi}^{\pi} [e^{i(\mathcal{K}+1)\theta} + e^{i(\mathcal{K}-1)\theta}] \\
    =\pi[\delta(\mathcal{K}+1) + \delta(\mathcal{K}-1)],
\end{split}
\end{equation}
where  $\mathcal{K}=m_j-m_i$ and the following expression has been used $\int e^{ip\vartheta}=2\pi\delta(p)$. Similarly, the matrix elements of the $y$-component of 
the dipole moment, $ D_{y,ij}^{(\zeta)}$, are proportional to 
\begin{equation}\label{eqn30}
    -i\pi[\delta(\mathcal{K}+1) - \delta(\mathcal{K}-1)].
\end{equation}
Then we can define $\Omega_1$ and $\Omega_2$ as follows 
\begin{equation}\label{eqn31}
    \begin{split}
        \Omega_x=\pi[\delta(m_i,m_j+1) + \delta(m_i,m_j-1)]\\
        \Omega_y=\pi[\delta(m_i,m_j+1) + \delta(m_i,m_j-1)].
    \end{split}
\end{equation}
%\bibliographystyle{apsrev4-1} % Tell bibtex which bibliography style to use

%\bibliography{TMDCBibliography}

% Tell bibtex which .bib file to use (this one is some example file in TexLive's file tree)

\end{document}